\documentclass[a4paper,11pt]{article}
\usepackage{pos}

\usepackage{xspace}
\usepackage{graphicx}
\usepackage[outdir=./]{epstopdf}

\newcommand {\meanpT}    {\ensuremath{\langle p_{\mathrm{T}} \kern-0.1em\rangle}\xspace}
\newcommand {\mean}[1]   {\ensuremath{\langle #1 \kern-0.1em\rangle}\xspace} 

\newcommand {\sqrts}     {\ensuremath{\sqrt{s}}\xspace}

\newcommand {\ee}        {\mbox{$\mathrm {e^+e^-}$}\xspace}
\newcommand {\ep}        {\mbox{$\mathrm {ep}$}\xspace}
\newcommand {\pp}        {\mbox{$\mathrm {pp}$}\xspace}

\newcommand {\MeanNpart} {\mbox{\ensuremath{< \kern-0.15em N_{part} \kern-0.15em >}}}

\newcommand {\mass}     {\mbox{\rm MeV$\kern-0.15em /\kern-0.12em c^2$}}
\newcommand {\tev}      {\mbox{${\rm TeV}$}\xspace}

\newcommand {\mmom}     {\mbox{\rm MeV$\kern-0.15em /\kern-0.12em c$}}
\newcommand {\gmom}     {\mbox{\rm GeV$\kern-0.15em /\kern-0.12em c$}}
\newcommand {\mmass}    {\mbox{\rm MeV$\kern-0.15em /\kern-0.12em c^2$}}
\newcommand {\gmass}    {\mbox{\rm GeV$\kern-0.15em /\kern-0.12em c^2$}}

\newcommand {\dg}       {\mbox{$\kern+0.1em ^\circ$}}

\newcommand{\gevc}{\ensuremath{\mathrm{GeV}/c}\xspace}

\newcommand{\pt}{\ensuremath{p_{\rm T}}\xspace}

\newcommand{\DZero}     {\ensuremath{\mathrm {D^0}}\xspace}

\newcommand{\BZero}     {\ensuremath{\mathrm {B^0}}\xspace}

\newcommand{\Bs}     {\ensuremath{\mathrm {B^0_s}}\xspace}

\newcommand{\Lb}{\ensuremath{\rm {\Lambda_b^{0}}}\xspace}
\newcommand{\Xic}         {\ensuremath{\mathrm {\Xi_{c}^{0,+}}}\xspace}

\newcommand{\lambdac}     {\ensuremath{\mathrm {\Lambda_{c}^{+}}}\xspace}

\newcommand{\rmLambdas}         {\ensuremath{\mathrm {\Lambda \kern-0.2em + \kern-0.2em \overline{\Lambda}}}\xspace}

\newcommand{\Dzero}{\ensuremath{\mathrm {D^0}}\xspace}

\newcommand{\Ds}{\ensuremath{\rm D_s^+}\xspace}

\newcommand{\Lcplus}{\lambdac}
\newcommand{\Lc}         {\Lcplus}
\newcommand{\LcD} {\ensuremath{\lambdac/\Dzero}\xspace}

\title{Heavy-flavour production at the LHC}

\author*[a,1]{Jaime Norman }

\affiliation[a]{University of Liverpool,\\
  Oliver Lodge Laboratory, Oxford St, Liverpool, L69 7ZE, UK}

\note{on behalf of the ALICE, ATLAS, CMS and LHCb collaborations}

\emailAdd{jaime.norman@cern.ch}

\abstract{Heavy flavour production measurements in \pp collisions are a crucial test of QCD. The LHC experiments ALICE, ATLAS, CMS and LHCb, provide complementary abilities to measure many aspects of heavy-flavour production. This contribution summarises recent LHC measurements within this topic.}

\FullConference{%
  The Tenth Annual Conference on Large Hadron Collider Physics - LHCP2022\\
  16-20 May 2022\\
  online
}

\begin{document}
\maketitle

Heavy-flavour (charm or beauty) hadron production in proton-proton collisions can be described with the factorisation approach as a convolution of the heavy quark partonic cross section, the parton distribution functions (PDFs) of the proton, and the fragmentation function (FF) of the heavy-flavour quark into a given hadron.
The hard parton cross sections are calculated with perturbative QCD techniques, where the large mass of the heavy quark sets the hard scale meaning production can be calculated over all $\pt$. The PDFs and FFs on the other hand must be determined through measurements. At the LHC, the quark and gluon densities in the proton means multiple hard-parton scattering becomes more relevant, and the underlying event activity becomes much larger, which may affect the fragmentation and hadronisation of quarks into hadrons. 
Measurements of heavy-flavour hadrons in general constitute some of the most important tests of QCD over many length scales at hadron colliders.

Heavy-flavour meson and quarkonia production has been studied extensively at the LHC. Measurements of the prompt~\cite{LHCb:2015swx,CMS:2021lab} and non-prompt (from beauty decays)~\cite{ALICE:2021mgk} production of D mesons agree well with QCD calculations~\cite{Kniehl:2012ti,Cacciari:1998it}, indicating that heavy-flavour meson production is well understood. The measurement of strange to non-strange beauty meson production has also been measured from non-prompt D mesons with ALICE~\cite{ALICE:2021mgk}, and a combined analysis of $\mathrm{B}$ meson decays from LHCb~\cite{LHCb:2021qbv}. The LHCb measurement hints at a slight increase in the production ratio of strange to non-strange beauty meson with collision energy. This is also a crucial measurement for reducing the uncertainty in many $\mathrm{B_s^0}$ branching fractions, which are a dominant source of uncertainty in many searches for new physics.
Prompt $J/\psi$ production has been measured by ALICE~\cite{ALICE:2021edd}, ATLAS~\cite{ATLAS-CONF-2019-047},  CMS~\cite{CMS:2017exb} and LHCb~\cite{LHCb:2021pyk}, 
and non-prompt $J/\psi$ production is also measured by ALICE~\cite{ALICE:2021edd},  ATLAS~\cite{ATLAS-CONF-2019-047} and CMS~\cite{CMS:2017exb}.

\begin{figure}[hbt]
	\centering
	\includegraphics[width=.39\textwidth]{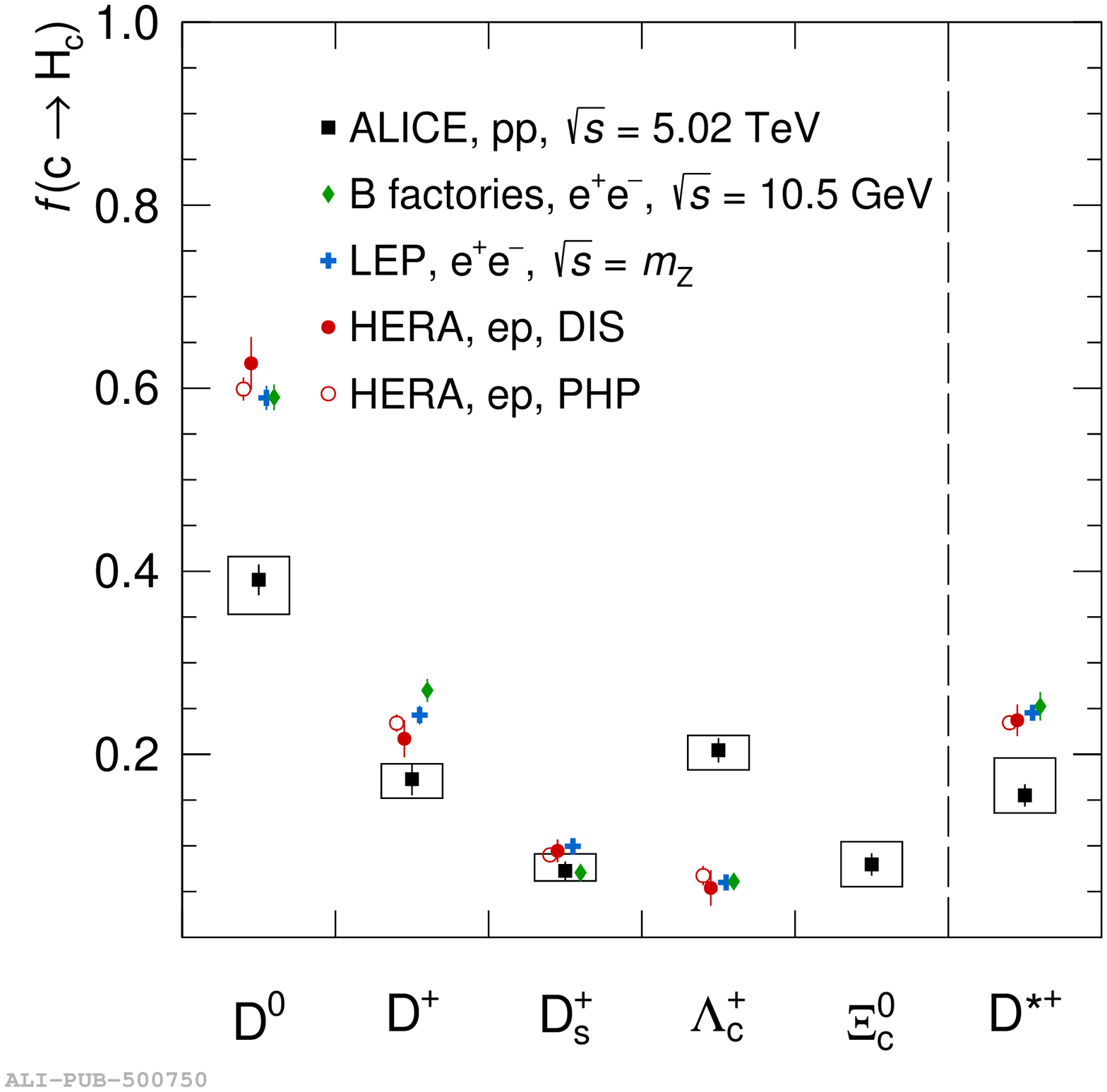}
	\includegraphics[width=.60\textwidth]{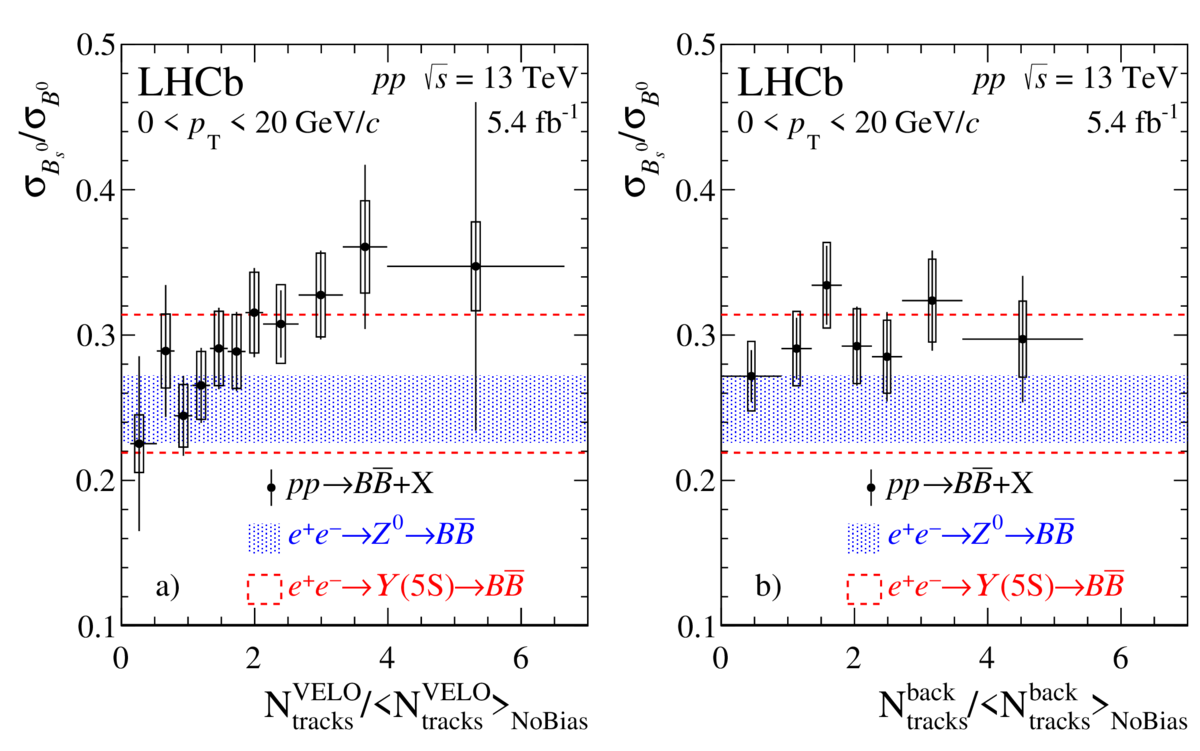}
	\caption[]{Left: Charm quark fragmentation fractions into charm hadrons in \pp collisions at $\sqrts = 5.02~\tev$, compared with measurements at LEP and B factories, and ep collisions~\cite{ALICE:2021dhb}. Right: The production ratios of strange to non-strange B mesons $\Bs / \mathrm{B^0}$ as a function of the event multiplicity measured either in the direction of the beauty meson (left) or in the opposite direction (right)~\cite{LHCb:2022syj}.}  
	\label{fig:FF}
\end{figure}

Comprehensive studies of heavy-flavour baryon production have recently been performed at the LHC, and currently heavy-flavour baryon production is less well understood than heavy-flavour meson production.  $\Lc$ production has been measured in \pp collisions by ALICE~\cite{ALICE:2021rzj,ALICE:2020wfu,ALICE:2020wla} and CMS~\cite{CMS:2019uws}, and is significantly underestimated by QCD calculations. The baryon-to-meson production ratio $\LcD$ is significantly larger than the same measurements in $\ee$ and $\ep$ collisions, by up to a factor of 5 at low $\pt$, and exhibits a strong $\pt$ dependence. Recently the baryon-to-meson ratios $\Xic / \DZero$~\cite{ALICE:2021psx,ALICE:2021bli}, $\Sigma_\mathrm{c}^{++,+} / \DZero$~\cite{ALICE:2021rzj} and $\Omega_\mathrm{c}^0 / \DZero$~\cite{ALICE:2022cop} have also been measured by ALICE to be significantly underestimated by predictions utilising fragmentation parameterisations based on measurements in $\ee$ and $\ep$ collisions. The fragmentation fractions of charm quarks into charmed hadrons have been measured for the first time in \pp collisions~\cite{ALICE:2021dhb}, which are shown in figure \ref{fig:FF}. The charmed hadron fragmentation fractions differ significantly from measurements in $\ee$ and $\ep$ collisions, indicating the assumption of universal, independent parton-to-hadron fragmentation across collision systems is not sufficient to describe charmed hadron production in \pp collisions at the LHC. 
Possible explanations of this difference include colour reconnection between independent partons~\cite{Christiansen:2015yqa}, quark coalescence~\cite{Minissale:2020bif,Song:2018tpv}, or enhanced baryon production originating from the decay of as-yet-undiscovered charm baryon states~\cite{He:2019tik}, which describe baryon-to-meson ratios better, though still underpredict the $\Omega_\mathrm{c}^0 / \DZero$ and $\Xic / \DZero$ ratios.
The relative production of beauty baryons and mesons has also been measured by LHCb~\cite{LHCb:2019fns}, where a similar enhancement of the ratio $\Lb / (\BZero + B^+$) is measured at low $\pt$. A measurement of the ratio of non-prompt $\Lc$ baryons and $\DZero$ mesons was made as a function of $\pt$ by ALICE which aims to indirectly probe beauty quark fragmentation. This measurement was compared to predictions utilising quark production with FONLL, and hadronisation/hadron decays with PYTHIA - the predictions require the fragmentation fractions measured by LHCb to describe the data, and the measurement is underestimated if utilising fragmentation fractions measured in $\ee$ collisions. These measurements also suggest that fragmentation fractions of beauty baryons are not universal in different collision systems. 
The production cross section measurements of different charmed hadrons has allowed for precise measurement of the total $c\bar{c}$~\cite{ALICE:2021dhb} and $b\bar{b}$~\cite{ALICE:2021edd} production cross sections.

Heavy-flavour hadron production has been studied as a function of the event multiplicity. The strange to non-strange production ratio of beauty mesons $\mathrm{B^0_s} / \mathrm{B^0}$ measured by LHCb in \pp collisions~\cite{LHCb:2022syj} is shown in figure \ref{fig:FF} (right). When measuring the multiplicity in the same rapidity region as the beauty hadron (with the VELO tracker) there is evidence at a level of $3.4\sigma$ that the ratio $\mathrm{B^0_s / B^0}$ increases with multiplicity. When measuring the multiplicity in the opposite rapidity direction as the beauty hadron, the ratio is instead independent of this multiplicity, indicating the enhancement is due to the local event multiplicity. 
The charmed baryon-to-meson ratio measured by ALICE~\cite{ALICE:2021npz} also displays a significant enhancement in the region $2 < \pt < 12~\gevc$ at high multiplicity compared to low multiplicity.  This multiplicity dependence is described well by PYTHIA when including colour reconnection mechanisms beyond the leading colour approximation. The charmed strange to non-strange ration $\Ds / \DZero$ is instead independent of multiplicity within the current experimental uncertainties~\cite{ALICE:2021npz}.

\begin{figure}[hbt]
	\centering
	\includegraphics[width=.39\textwidth]{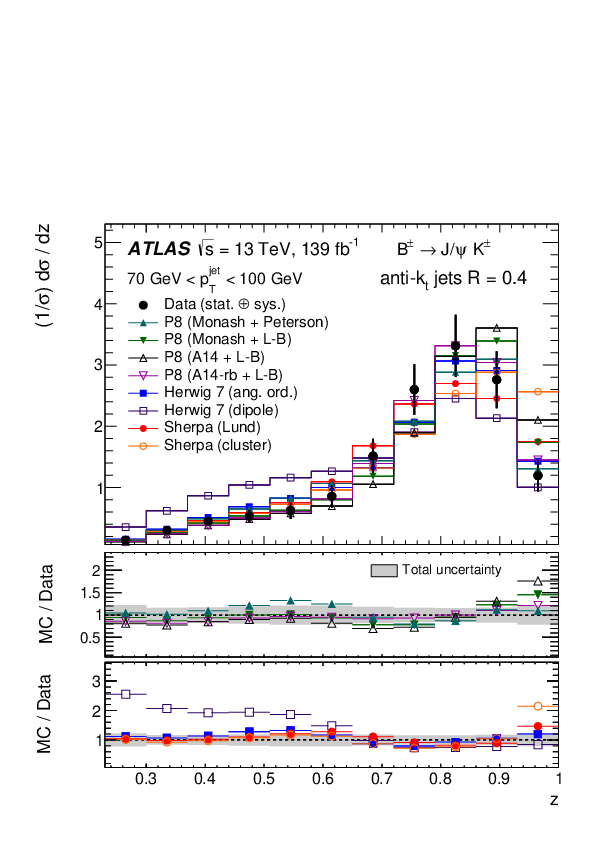}
	\includegraphics[width=.45\textwidth]{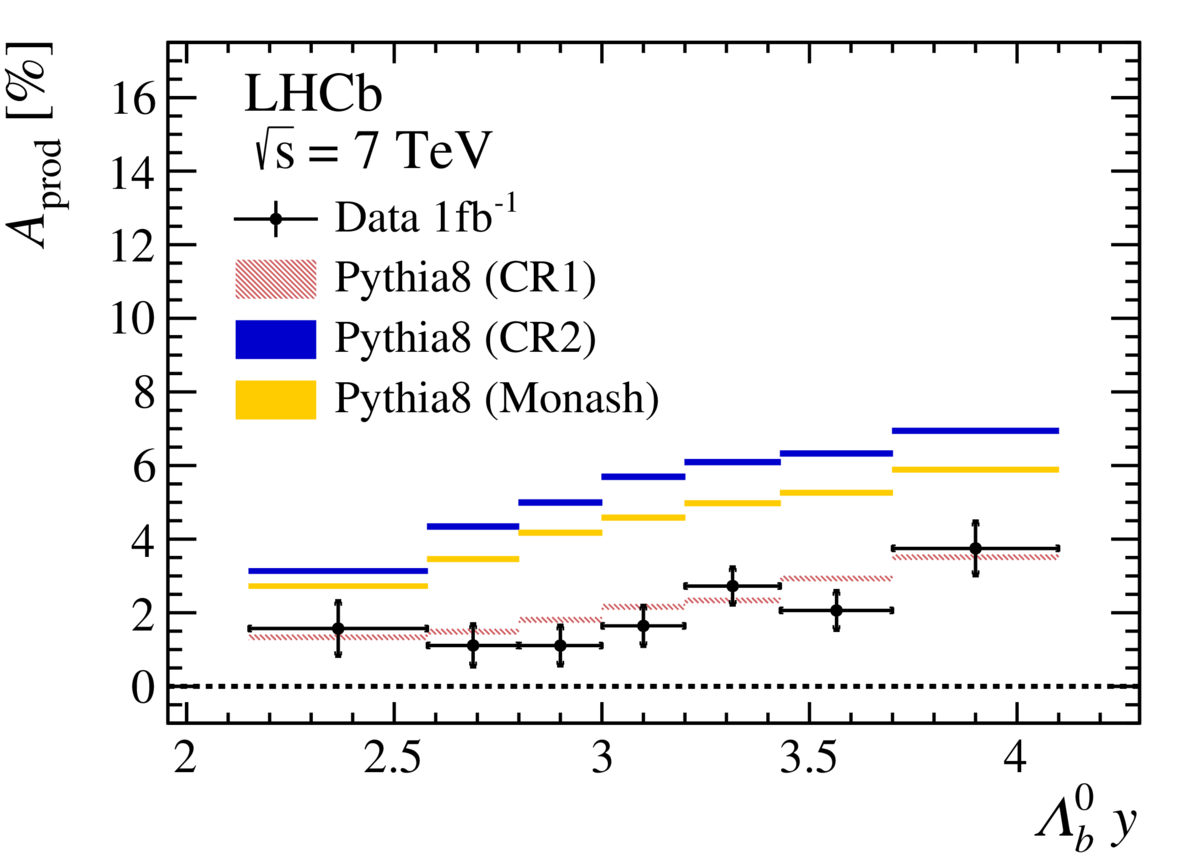}
	\caption[]{Left: The distribution of the longitudinal profile $z$ of $\mathrm{B}^\pm$ mesons in jets~\cite{ATLAS:2021agf}. Right: the production asymmetry of $\Lb$ baryons~\cite{LHCb:2021xyh}.}  
	\label{fig:jet}
\end{figure}

Fragmentation dynamics of heavy quarks are probed with measurements of heavy flavour hadrons within jets. Figure \ref{fig:jet} (left) shows a measurement of $\mathrm{B}^{\pm}$ mesons within jets by ATLAS~\cite{ATLAS:2021agf} - in particular, the relative momentum of the jet that is carried by the hadron in the direction of the jet, $z = \frac{\vec{p}_{B,D} \cdot \vec{p}_j}{|\vec{p}_j|}$. The transverse momentum profile is also reported in ~\cite{ATLAS:2021agf}. 
ALICE measured the $z$ distribution of D mesons~\cite{ALICE:2022mur} at  $\sqrt{s} = 5.02~\tev$ and $\sqrt{s} = 13~\tev$. Comparisons to Monte Carlo event generators were made in all cases, which helps constrain different approaches to fragmentation in these generators.

The first observation of $b$-hadron production asymmetry (i.e., asymmetrical production of a beauty hadron compared to its anti-particle, measured using $A_{prod} = \frac{\sigma(pp\rightarrow\Lambda_b^0 Y) - \sigma(pp\rightarrow\bar{\Lambda^0_b} Y ) }{\sigma(pp\rightarrow\Lambda_b^0 Y) + \sigma(pp\rightarrow\bar{\Lambda^0_b} Y ) }$) was reported by LHCb~\cite{LHCb:2021xyh}. Figure \ref{fig:jet} shows $A_{prod}$ as a function of $\Lb$ rapidity at $\sqrts = 7~\tev$. Combining the measurement at $\sqrts = 7~\tev$ and $8~\tev$, the results are incompatible with symmetric production with a significance of 5.8 standard deviations, assuming no CP violation in the decay. There is also evidence at the level of $4\sigma$ of an increase in production asymmetry with rapidity. Comparisons from MC generators show that the results agree with PYTHIA when an improved colour reconnection model is included.

CMS reported the observation of simultaneous production of three $J/\psi$ meson particles~\cite{CMS:2021qsn}. In this analysis, five events are found to be consistent with triple-$J/\psi$ production, with a statistical significance relative to the background-only expectation of 5 standard deviations. The measured cross section is $\sigma(\pp \rightarrow J/\psi J/\psi J/\psi X) = 272 \substack{+141 \\ -104} \mathrm{(stat.)} \pm 17 \mathrm{(syst.)}$ fb. This cross section is consistent with theoretical expectation of production via primarily double-parton scattering and triple-parton scattering (under the simplest assumption of factorization of multiple hard-scattering probabilities in terms of SPS cross sections). This measurement is the first observation of the simultaneous production of three heavy particles, and represents an important step in the quest to constrain the proton PDF.

In summary, heavy-flavour production at the LHC constitutes a stringent test of QCD, and Run 2 of the LHC has allowed for a diverse range of heavy-flavour production measurements. Run 3 has just begun after significant detector upgrade programmes for all 4 LHC experiments during Long Shutdown 2, which will provide unprecedented accuracy for future heavy-flavour production measurements.


\clearpage

\bibliographystyle{JHEP}
\bibliography{include/bibfile}


\end{document}